\documentclass[twocolumn]{aastex631}

\shorttitle{A pulsating Mira donor in Scutum X-1}
\shortauthors{K. De et al.}

\graphicspath{{./}{figures/}}

\begin{document}

\title{A massive AGB donor in Scutum X-1: Identification of the first Mira variable in an X-ray binary}

\author[0000-0002-0786-7307]{Kishalay De}
\altaffiliation{NASA Einstein Fellow}
\affiliation{MIT-Kavli Institute for Astrophysics and Space Research, 77 Massachusetts Ave., Cambridge, MA 02139, USA}

\author{Deepto Chakrabarty}
\affiliation{MIT-Kavli Institute for Astrophysics and Space Research, 77 Massachusetts Ave., Cambridge, MA 02139, USA}

\author[0000-0002-4622-796X]{Roberto Soria}
\affiliation{College of Astronomy and Space Sciences, University of the Chinese Academy of Sciences, Beijing 100049, People's Republic of China}
\affiliation{Sydney Institute for Astronomy, School of Physics A28, The University of Sydney, Sydney, NSW 2006, Australia}

\author[0000-0003-1412-2028]{Michael C. B. Ashley}
\affil{School of Physics, University of New South Wales, Sydney NSW 2052, Australia}

\author{Charlie Conroy}
\affil{Harvard-Smithsonian Center for Astrophysics, Cambridge, MA 02138, USA}

\author[0000-0001-9315-8437]{Matthew J. Hankins}
\affil{Arkansas Tech University, Russellville, AR 72801, USA}

\author[0000-0002-5619-4938]{Mansi M. Kasliwal}
\affil{Cahill Center for Astrophysics, California Institute of Technology, 1200 E. California Blvd. Pasadena, CA 91125, USA.}

\author{Ryan M. Lau}
\affil{Institute of Space \& Astronautical Science, Japan Aerospace Exploration  Agency,  3-1-1 Yoshinodai,  Chuo-ku,  Sagamihara, Kanagawa 252-5210, Japan}

\author{Anna M. Moore}
\affil{Research School of Astronomy and Astrophysics, Australian National University, Canberra, ACT 2611, Australia}

\author{Robert Simcoe}
\affiliation{MIT-Kavli Institute for Astrophysics and Space Research, 77 Massachusetts Ave., Cambridge, MA 02139, USA}

\author{Jamie Soon}
\affil{Research School of Astronomy and Astrophysics, Australian National University, Canberra, ACT 2611, Australia}

\author[0000-0001-9304-6718]{Tony Travouillon}
\affil{Research School of Astronomy and Astrophysics, Australian National University, Canberra, ACT 2611, Australia}

\correspondingauthor{Kishalay De}
\email{kde1@mit.edu}

\begin{abstract}
The symbiotic X-ray binary Sct X-1 was suggested as the first known neutron star accreting from a red supergiant companion. Although known for nearly 50 years, detailed characterization of the donor remains lacking, particularly due to the extremely high reddening towards the source ($A_V\gtrsim25$\,mag). Here, we present i) improved localization of the counterpart using {\it Gaia} and {\it Chandra} observations, ii) the first broadband infrared spectrum ($\approx1-5\,\mu$m; $R\approx 2000$) obtained with SpeX on the NASA Infrared Telescope Facility and iii) $J$-band light curve from the Palomar Gattini-IR survey. The infrared spectrum is characterized by i) deep water absorption features (H$_2$O index $\approx 40$\%), ii) strong TiO, VO and CO features, and iii) weak/absent CN lines. We show that these features are inconsistent with known red supergiants, but suggest a M8-9 III type O-rich Mira donor star. We report the discovery of large amplitude ($\Delta J\approx3.5$\,mag) periodic photometric variability suggesting a pulsation period of $621\pm36\,{\rm(systematic)}\pm8\,{\rm(statistical)}$\,days, which we use to constrain the donor to be a relatively luminous Mira ($M_K=-8.6\pm0.3$\,mag) at a distance of $3.6^{+0.8}_{-0.7}$\,kpc. Comparing these characteristics to recent models, we find the donor to be consistent with a $\approx 3-5$\,M$_\odot$ star at an age of $\approx 0.1-0.3$\,Gyr. Together, we show that Sct\,X-1 was previously mis-classified as an evolved High Mass X-ray Binary; instead it is an intermediate mass system with the first confirmed Mira donor in an X-ray binary. We discuss the implications of Mira donors in symbiotic X-ray binaries, and highlight the potential of wide-field infrared time domain surveys and broadband infrared spectroscopy to unveil their demographics.

\end{abstract}


\section{Introduction} 

Scutum X-1 was identified  from a sounding rocket experiment nearly 50 years ago \citep{Hill1974}, as an unusual and bright X-ray source exhibiting the strongest absorption ever detected at the time. The source has been subsequently detected by several X-ray missions (see \citealt{Kaplan2007} for a review), and has been noted for exhibiting a conspicuous $\approx 112$\,s X-ray pulsation arising from a neutron star (NS) accretor \citep{Koyama1991}. \citet{Kaplan2007} presented the first accurate (to $\approx 1$\arcsec)  X-ray localization using {\it XMM Newton} observations, which positioned the X-ray source to be within $0.2$\arcsec of a bright infrared counterpart ($K_s \approx 6.5$\,mag) in the 2MASS catalog. The counterpart was reported to be nearly invisible in the optical bands ($r>25.2$\,mag, $i \approx 23.6$\,mag; \citealt{Kaplan2007}). Based on the extremely red colors ($J-K \approx 5.5$\,mag), $HK$-band spectroscopy, and observed X-ray absorption, they suggested the counterpart to likely be a late-type supergiant with a two dimensional classification of $\approx$M\,0-1 Ia-Iab. The source was thus classified as a symbiotic X-ray binary (SyXRB; \citealt{Masetti2006}) at a distance of $4-10$\,kpc, and behind an extinction column of $A_V \gtrsim 25$\,mag.

The suggested classification makes Sct\,X-1 the first known candidate for a NS accreting from the wind of a red supergiant (RSG) in a Galactic X-ray binary\footnote{The \texttt{SIMBAD} database lists Sct X-1 as a High Mass X-ray Binary}, and an evolved member of the high mass X-ray binary (HMXB) population. The source still remains one of only three such known systems -- the other two sources being the Galactic center X-ray binary CXO\,174528.79-290942.8 \citep{Gottlieb2020}, and 4U\,1954+31 for which the donor classification was only recently revised from a M4-5 III giant \citep{Masetti2006} to a M4 I supergiant \citep{Hinkle2020}. Unlike the more common B-supergiant HMXBs \citep{MartinezNunez2017}, RSG HMXBs are rare due to their short lifetimes, but are of interest as ubiquitous evolutionary phases in the path from massive binaries to double compact objects \citep{Mondal2020}.

As noted in \citet{Kaplan2007}, the spectrum used to classify the donor in Sct\,X-1 has uncertain flux calibration as well as limited wavelength coverage. In addition, the high reddening towards the source precludes any attempts at optical spectroscopy or photometry to constrain the donor type via spectral features or variability. Concomitantly, the {\it Gaia} catalog does not provide a parallax to constrain its distance and luminosity. The emergence of long term infrared time domain surveys combined with moderate aperture infrared spectroscopy offers new opportunities to re-visit the demographics of obscured Galactic X-ray sources such as Sct\,X-1. 

In this paper, we present the first broadband infrared spectrum and near-infrared (NIR) light curve of the infrared counterpart of Sct\,X-1. Using the  observed  strong molecular absorption features and large amplitude photometric variability, we show that Sct\,X-1 has been long mis-classified as a RSG X-ray binary, and instead contains a very late-type pulsating Mira donor star. We present the observational data in Section \ref{sec:obs}. Section \ref{sec:anal} presents an analysis of the infrared spectral features and photometric variability to constrain the donor type, distance and extinction. We discuss the implications of these results, and summarize our findings in Section \ref{sec:disc}.

\section{Observations}
\label{sec:obs}

\subsection{Infrared photometry}

\begin{figure}
    \centering
    \includegraphics[width=\columnwidth]{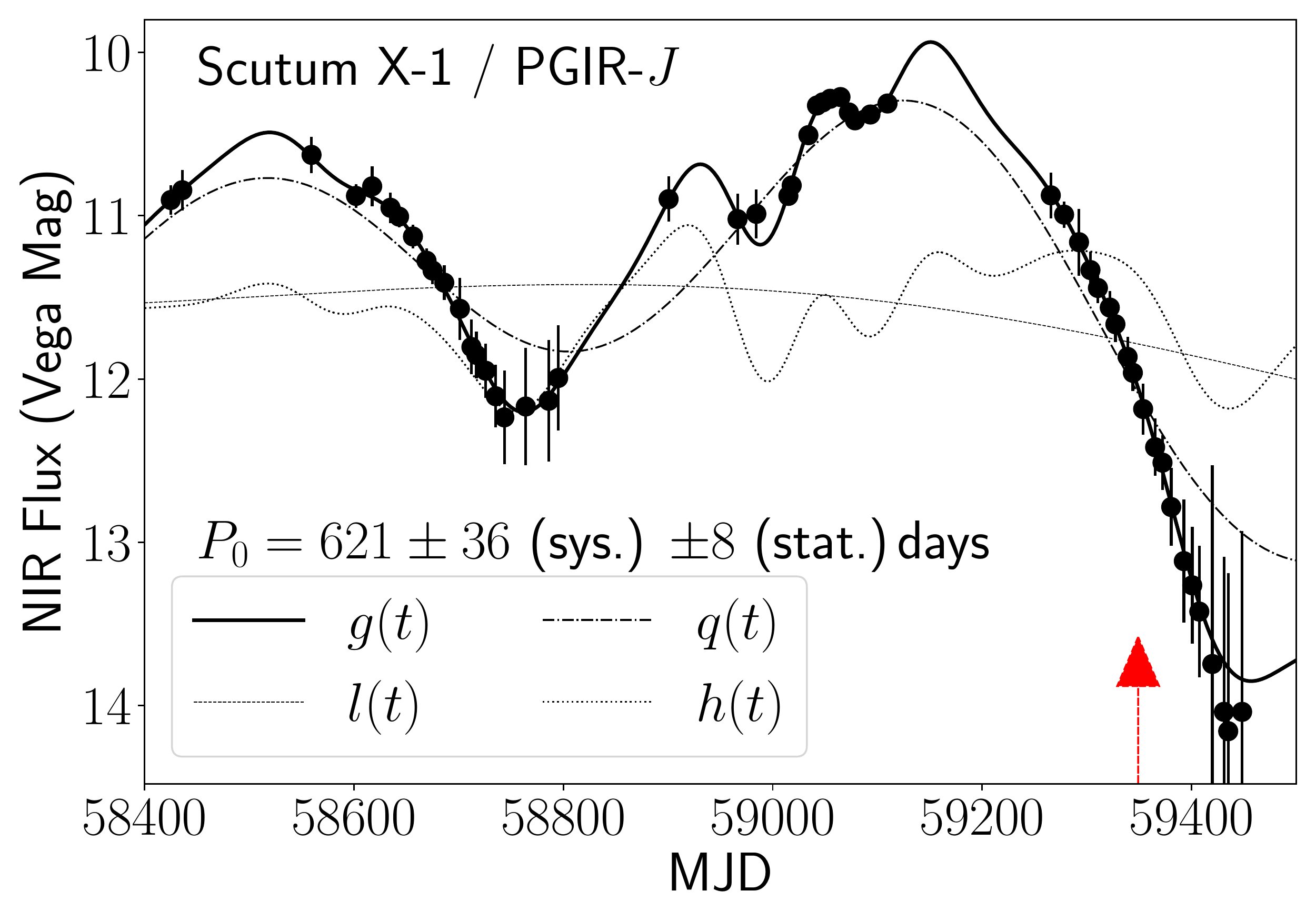}
    \caption{PGIR $J$-band light curve of 2MASSJ1835. The black points denote the observed $J$-band photometry. The black solid line shows the best-fit semi-parametric Gaussian process regression model ($g(t)$; see text). The black dashed, dot-dashed and dotted lines denote the contribution from the long-term trend component ($l(t)$), primary periodic component ($q(t)$) and high frequency variability component ($h(t)$) respectively. The best-fit period and its uncertainty is shown. The red arrow shows the SpeX observation epoch.}
    \label{fig:pgirlc}
\end{figure}

\citet{Kaplan2007} showed that the source 2MASS\,J18352582-0736501 (hereafter 2MASSJ1835) was very likely (chance probability $\sim 2 \times 10^{-6}$) the infrared counterpart of Sct\,X-1, with a 2MASS magnitude of $J\approx 12$\,mag. The sky location was monitored as a part of regular survey operations of the Palomar Gattini-IR NIR time domain survey \citep{De2020a, Moore2019}. Palomar Gattini-IR is a wide-field ($\approx 25$\,sq. deg.) and shallow $J$-band survey (reaching $J\approx 13$\,mag and $J\approx 15$\,mag\footnote{Vega magnitudes calibrated to the 2MASS photometric system are used throughout this paper} in and outside the Galactic plane, respectively), monitoring the entire visible sky from Palomar Observatory using a 30\,cm telescope with a cadence of $\approx 2$\,nights. The light curve was obtained by performing forced aperture photometry at the location of the infrared counterpart, with an aperture size of $8.3$\arcsec. The light curve is shown in Figure \ref{fig:pgirlc}, and was binned over $\approx 7$\,day intervals to improve the signal-to-noise ratio.

\subsection{Infrared spectroscopy}

We obtained an infrared spectrum of 2MASSJ1835 using SpeX on the NASA Infrared Telescope Facility (\citealt{Rayner2003}; Program 2021A083, PI: De) on UT 2021-05-16. The data were obtained with a 0.3\arcsec\ slit, in both the \textit{SXD} and \textit{LXD\_long} modes over multiple dithered exposures amounting to a total exposure time of $\approx 720$\,s and $\approx 300$\,s respectively. The standard stars HIP\,90967 and HIP\,87108 were used for telluric and flux calibration in the \textit{SXD} and \textit{LXD\_long} modes respectively. The combined spectrum covers the full wavelength range of the instrument from $\approx 0.8$ to $\approx 5.0\,\mu$m with a resolution of $R\approx 2000$. The data were reduced using the \texttt{spextool} pipeline \citep{Cushing2004}, while telluric and flux calibration were performed using the \texttt{xtellcor} package \citep{Vacca2003}. Due to the high reddening, the source is not detected at $< 1.0\,\mu$m in the final spectrum shown in Figure \ref{fig:irtfspec} and Figure \ref{fig:jspec} (observation epoch is shown with a red arrow in Figure \ref{fig:pgirlc}). The average signal-to-noise ratio is $\approx 300$ in $K$-band, $\approx 200$ in $H$-band and $\approx 70$ in $J$-band.

\subsection{X-ray observations}

The {\it Chandra X-ray Observatory} observed Sct X-1 on 2020 November 9 (ObsID 22417). The exposure started at MJD 59162.08 for a live time of 27.3\,ks. The live time was 27.3 ks. The source was located at the aim point on chip S3 of the Advanced CCD Imaging Spectrometer (ACIS) array. We downloaded the data from the public archive and processed them with the Chandra Interactive Analysis of Observations ({\sc ciao}) software version 4.12 \citep{Fruscione2006}, with calibration database version 4.9.1. We re-built a level-2 event file with the task {\it {chandra\_repro}}, and created images in different energy bands with {\it dmcopy}. We built background-subtracted spectra and associated response and ancillary response files with {\it specextract}. We regrouped the spectrum to at least 1 count per bin, with the {\sc ftools} task {\it grppha}, and fitted it with {\sc{xspec}} \citep{Arnaud1996} version 12.11.0, using the Cash statistics \citep{Cash1979}.

\section{Analysis}
\label{sec:anal}

\subsection{Improved localization and recent X-ray flux}

2MASSJ1835 is detected as a faint source in {\it Gaia} EDR3 \citep{GaiaEDR3} at the J2000 position $\alpha =$ 18:35:25.82, $\delta =$ -07:36:50.4 at a {\it Gaia} magnitude of $G \approx 20.5$. We used the {\sc ciao} analysis tools within the {\sc ds9} imaging package to determine the centroid of the point-like X-ray source in the 0.3--7 keV band {\it Chandra} data. The X-ray source centroid is measured to be $\alpha =$ 18:35:25.82, $\delta =$ -07:36:50.6. We cannot improve on the default {\it Chandra} astrometry because there are no other X-ray sources in the field of view. Therefore, the position contains a default 90\% uncertainty radius of $0.8$\arcsec, improving on the $\approx 1.5$\,\arcsec (root mean square; \citealt{Kirsch2004}) positional uncertainty of the previous {\it XMM Newton} observations. The position is coincident within $\approx 0.2$\,arcsec of the {\it Gaia} source, and confirms the association proposed by \citet{Kaplan2007} with the excellent {\it Chandra} and {\it Gaia} astrometry. 

Given the small number of counts ($\approx$170 net counts), we used only two simple models for spectral fitting to estimate the recent X-ray flux. We tried power-law and bremsstrahlung models given the current low X-ray luminosity of the source. In both cases, we included a free photo-electric absorption component, modelled with \texttt{tbabs}, with the abundances of \citet{Wilms2000}. A power-law model with photon index $\Gamma = 1.9^{+0.9}_{-0.8}$ (90\% confidence limit), and a bremsstrahlung model with $kT > 3.5$ keV, provide equally good fits (C statistics of 126.2/118 d.o.f, and 126.0/118 d.o.f., respectively). The column density of the neutral absorber is $N_{\rm H} = 5.6^{+2.4}_{-1.9}\,10^{22}$ cm$^{-2}$ for the power-law model, and $N_{\rm H} = 5.2^{+1.8}_{-1.5}\,10^{22}$ cm$^{-2}$ for the bremsstrahlung model. The absorbed 0.3--10 keV flux is $\approx$2.0 $\times 10^{-13}$ erg cm$^{-1}$ s$^{-1}$ in both cases. For the power-law model, this corresponds to a 0.3--10 keV unabsorbed flux of $6.3^{+15.0}_{-2.6} \times 10^{-13}$ erg cm$^{-2}$ s$^{-1}$; for the bremsstrahlung model, of $4.6^{+2.1}_{-1.0} \times 10^{-13}$ erg cm$^{-2}$ s$^{-1}$. The X-ray source has thus faded in flux by a factor of $\sim20-30$ since the 2004 {\it XMM-Newton} observations \citep{Kaplan2007}, $\sim 100\times$ compared with the 1987 GINGA observations \citep{Koyama1991} and a few $\sim 1000\times$ since the discovery in the 1970s \citep{Hill1974}.

\subsection{Infrared spectrum}
\label{sec:spectrum}

\begin{figure*}[!htp]
    \centering
    \includegraphics[width=\textwidth]{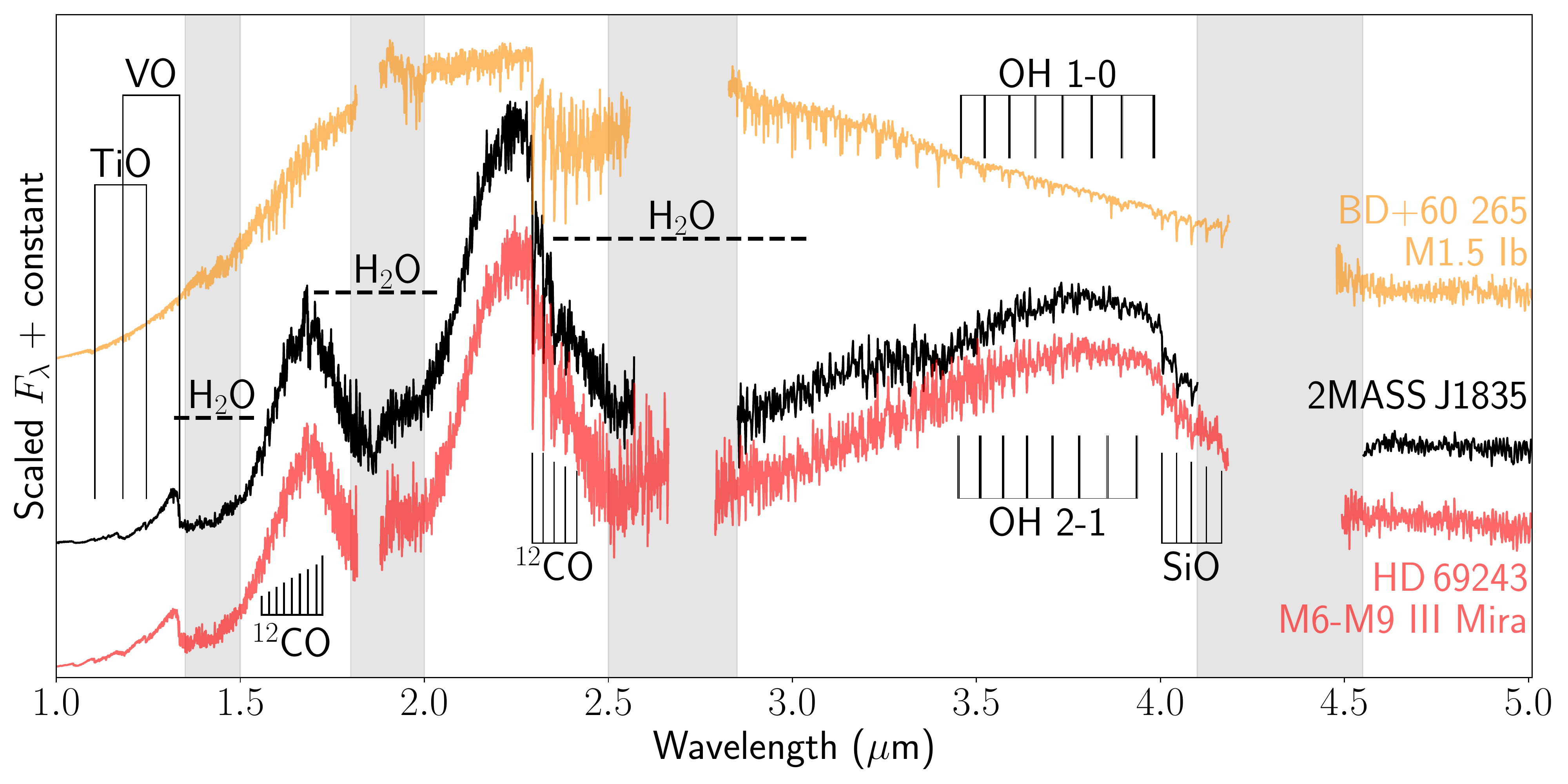}
    \caption{Comparison of the infrared spectrum of 2MASSJ1835 to that of a RSG (of the favored spectral type in \citealt{Kaplan2007}) and a late-type Mira variable (inferred from our analysis) in the IRTF Spectral Library \citep{Rayner2009}. Prominent absorption features are marked, while regions of low atmospheric transmission are shown in gray. The comparison spectra have been artificially reddened with $E(B-V) = 7.0$\,mag based on the estimate in \citealt{Kaplan2007} (see also Section \ref{sec:lcanal}). }

    \label{fig:irtfspec}
\end{figure*}

The broadband infrared spectrum of 2MASSJ1835 is shown in Figure \ref{fig:irtfspec}. The source exhibits a steep rising continuum from $\approx 1.0$ to $\approx 2.3\,\mu$m, consistent with the high reddening reported by \citet{Kaplan2007}. Superimposed on the rising continuum, the spectrum exhibits broad and deep absorption features around $\approx 1.4$, $\approx 1.9\,\mu$m and $\approx 2.7\,\mu$m, consistent with water absorption features seen in very late-type giants \citep{Kleinmann1986, Rayner2009}. At shorter wavelengths, the strongest absorption features are the bandheads of TiO and VO (in $J$-band), while we identify absorption features of the second $^{12}$CO overtone in $H$-band. The $K$-band is dominated by conspicuous sharp features of the $^{12}$CO first overtone series, while we detect features of the OH 1-0, 2-1 and SiO bandheads in $L$-band.

Although \citet{Kaplan2007} did not conclusively identify the spectral and luminosity class of 2MASSJ1835 they argued for a similarity with that of early M-type supergiants based on the detection of metal features and strong $^{12}$CO $K$-band features \citep{Ramirez1997, Comeron2004}. The $^{12}$CO line equivalent width (EW) is known to increase with decreasing temperature and specific gravity, making it particularly well suited to identifying cool supergiants \citep{Rayner2009}. However, very late type Asymptotic Giant Branch (AGB) stars, typically appearing as Mira variables, also exhibit similarly strong (and variable) $^{12}$CO features as supergiants \citep{Blum2003, Comeron2004}, making the previous classification unreliable.

\citet{Messineo2021} recently devised a classification method for giants, Miras and supergiants using a large sample of IRTF spectra, particularly focusing on the Mg\,I $1.71\,\mu$m to break the degeneracy between temperature and gravity. Based on this scheme, we measure a $^{12}$CO EW of $50.5 \pm 1.5$\,\AA\ and a Mg\,I $1.71\,\mu$m EW of $3.0 \pm 0.6$\,\AA. The values place the source squarely in the phase space occupied by only supergiant stars and Mira AGB stars (see their Figure 17), ruling out the case for a normal giant. Strong water absorption features\footnote{\citet{Kaplan2007} also noted possible water absorption in their $K$-band spectra but were inconclusive due to uncertain flux calibration.} are known to be seen only in large amplitude pulsating variables ($\delta V > 1.7$\,mag; \citealt{Lancon2000}), that are formed in the propagating shocks in their dense extended atmospheres \citep{Bessell1989}. Based on the H$_2$O absorption index defined in \citet{Blum2003}, we measure an absorption of $\approx 40$\%. The H$_2$O absorption index increases with decreasing effective temperature and decreasing luminosity \citep{Lancon1992,Comeron2004}; hence the large absorption clearly suggests this source to be a very late-type Mira variable since supergiants exhibit absorption indices of $\lesssim 6$\% \citep{Messineo2021}. Figure \ref{fig:irtfspec} shows a comparison of 2MASSJ1835 to that of the favored stellar type in \citet{Kaplan2007} and a late-type Mira variable, clearly demonstrating consistency with the latter.

\begin{figure}
    \centering
    \includegraphics[width=\columnwidth]{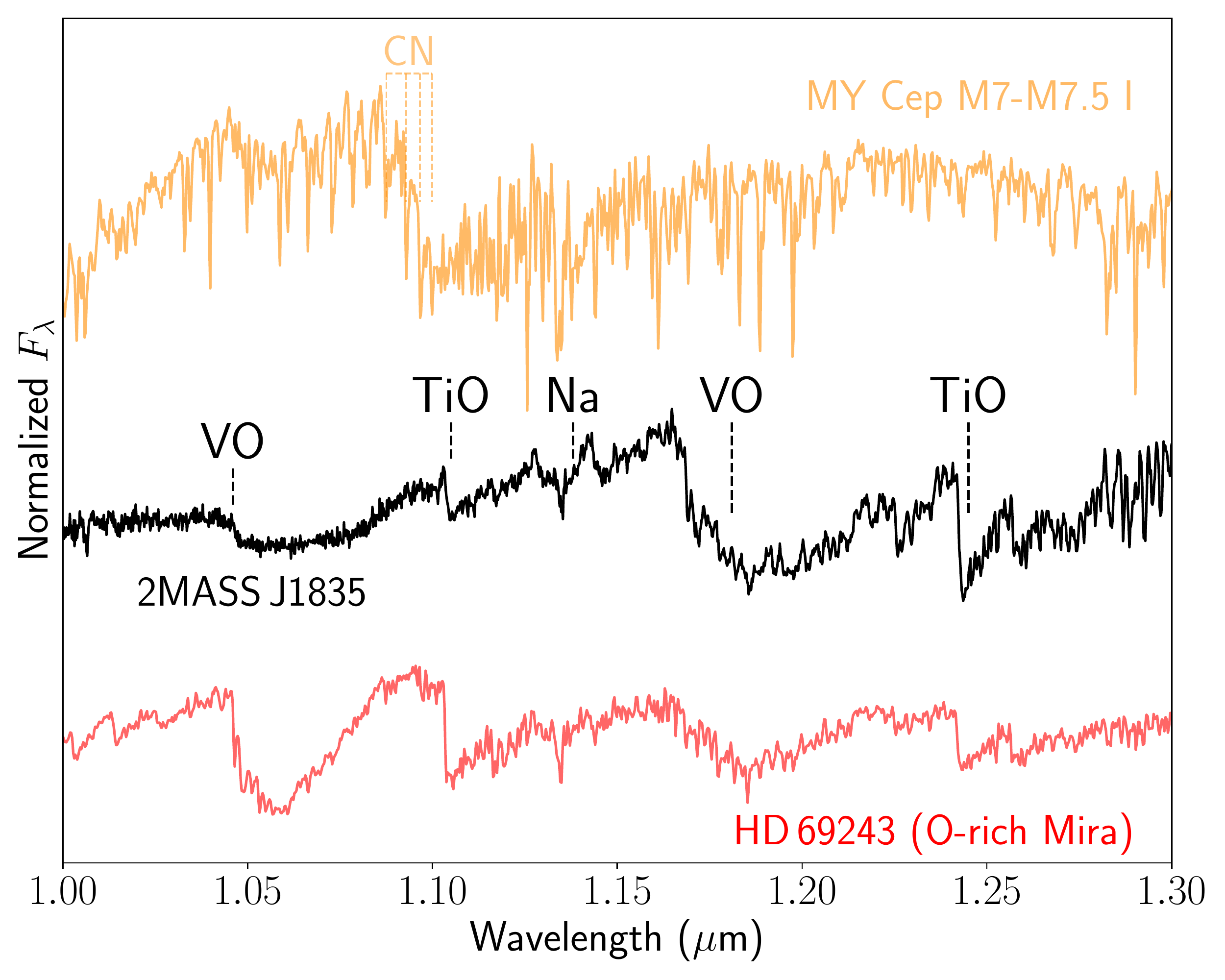}
    \caption{Comparison of the $J$-band spectrum of 2MASSJ1835 to that a similar late-type supergiant (MY Cep) and Mira variable (HD\,69243) in the IRTF spectral library. The spectra have been normalized by fitting the steep continuum with a low order polynomial.}
    \label{fig:jspec}
\end{figure}

\citet{Messineo2021} highlight additional differences between the NIR spectra of Miras and supergiants based on molecular features of VO, TiO and CN in $J$-band. Figure \ref{fig:jspec} shows the $J$-band spectrum of 2MASSJ1835, highlighting deep absorption bandheads of VO and TiO, compared to that of a late-type supergiant and a Mira variable. The TiO bandheads near 1.25\,$\mu$m are detected at spectral types M7 or later \citep{Wright2010}, while the VO features are known to become prominent at very low effective temperatures of $T_{\rm eff} \lesssim 3200$\,K (spectral type M6 or later; \citealt{Joyce1998}). In particular, the TiO bandhead at $\approx 1.1\,\mu$m is only observed in the spectra of O-rich Mira AGBs, while the VO feature at $\approx 1.05\,\mu$m is strong in O-rich Mira AGBs but weak in supergiants \citep{Messineo2021}. RSGs also ubiquitously show CN absorption bandheads near $\approx 1.09\,\mu$m since their atmospheres are rich in CN molecules, which are clearly seen in the spectrum of the RSG but are weak/absent in 2MASSJ1835. We measure the EW of the CN absorption features using the \texttt{J8 + J9 + J10} index defined in \citet{Messineo2021}, and measure a value of $0.03 \pm 1.07$\,\AA\ and consistent with the phase space only occupied by Mira AGBs (see Figure 5 in \citealt{Messineo2021}). Overall, we conclude that the molecular absorption features in 2MASSJ1835 suggest a revised classification of the donor star to be a M8-9 III type Oxygen-rich Mira variable.

\subsection{Infrared light curve}
\label{sec:lcanal}
\subsubsection{Periodicity analysis}

The well-sampled $J$-band light curve of 2MASSJ1835 is shown in Figure \ref{fig:pgirlc}. The light curve shows a clear periodic trend superimposed with smaller amplitude high frequency oscillations. Noting the strong evidence for water absorption in the counterpart spectrum (see Section \ref{sec:spectrum}), we interpret the variability as large amplitude pulsations of a Mira variable \citep{Kholopov1985}. Mira variables are known to be not strictly periodic, but characterized by both low frequency and high frequency variations associated with variations in circumstellar dust and supergranular convection \citep{Groenewegen2007, Iwanek2021}. We use the semi-parametric Gaussian process regression model proposed by \citet{He2016} to account for these variations, decomposing the light curve into four parts
\begin{equation}
    g(t) = m + l(t) + q(t) + h(t)
\end{equation}
where $m$ is the mean magnitude, $l(t)$ is the slowly variable mean magnitude, $q(t)$ is the primary periodic term, and $h(t)$ is the high frequency stochastic variability. The terms are modeled Gaussian processes with different kernels to account for effects such as changing amplitudes across cycles.

 We use the \texttt{R} code distributed by \citet{He2016} to derive a primary pulsation period from the log-likelihood periodogram and a best-fit $J$-band light curve model including all the variability terms. We estimate the statistical uncertainty in the period by simulating 100 realizations of the observed light curve by adding Gaussian noise scaled to the photometric uncertainties. In order to estimate the systematic uncertainty given the relatively short baseline ($\approx 1000$ days) of our light curve, we selected a two Mira variables (OGLE-BLG-LPV-180685 and GLE-LMC-LPV-00055) with similar light curves to Sct X-1 from the OGLE-III database \citep{Udalski2008}, scaled their well-known periods to 621 days, and then examined the scatter of periods found from about fifty 1000-day chunks uniformly distributed along the full lightcurves. The standard deviation of the periods was 28 days for OGLE-BLG-LPV-180685, and 36 days for OGLE-LMC-LPV-00055, and we thus take the larger as an estimate of the uncertainty. The best-fit light curve and period are shown in Figure \ref{fig:pgirlc}. The observed peak-to-peak amplitude of $\Delta J \approx 3.5$\,mag is consistent with the large amplitude NIR variability seen in long period symbiotic Miras \citep{Gromadzki2009}. We discuss the implications of the observed pulsation amplitude on the reported archival fluxes of 2MASS\,J1835 in Appendix \ref{sec:app_colors}.


\subsubsection{Distance and extinction}

Mira variables exhibit a period-luminosity-color relation over several decades in luminosity \citep{Feast1989, Wood2000}, and that is commonly used to estimate distances to symbiotic Miras \citep{Whitelock1987, Gromadzki2009}. Using the \citet{Whitelock2008} period-luminosity relation for O-rich Miras suggests an intrinsic absolute magnitude of $M_K = -8.6 \pm 0.3$\,mag ($M_{\rm bol} \approx -5.1$). Our inferred NIR luminosity is marginally fainter than the supergiant luminosity ($M_K \approx -9.4$) class assumed in \citet{Kaplan2007}. Using the period-color relation from \citet{Whitelock2000}, we estimate a intrinsic NIR color of $(J-K)_0 = 1.59 \pm 0.26$\,mag. The 2MASS $J$-magnitude of the source is consistent with the long term average observed in the PGIR light curve, and hence we use the 2MASS measurements to infer a $E(J-K) = 3.92 \pm 0.27$\,mag.

The estimated color excess corresponds to an optical extinction of $A_V = 21.8 \pm 1.5$\,mag ($A_K = 2.4 \pm 0.2$\,mag) using the same extinction law as in \citet{Kaplan2007}. The estimate corroborates the previous suggestion that Sct\,X-1 lies behind a large column of dust. The observed $K$-band magnitude and inferred extinction places the source at a distance of $3.6^{+0.8}_{-0.7}$\,kpc. The estimated distance is consistent that estimated by \citet{Kaplan2007}, due to the inferred $K$-band luminosity difference being compensated by the redder intrinsic color of the Mira. Applying the inferred extinction (we assume a \citealt{Cardelli1989} extinction law with $R_V = 3.1$) to the observed 2MASS photometry and fitting a Planck function, we derive a NIR color temperature of $\approx 2000 \pm 190$\,K, consistent with the effective temperatures of very late-type Mira variables \citep{Haniff1995, Feast1996}. Overall, the long pulsation period and large amplitude is consistent with a very late-type Mira variable as inferred from the spectral analysis \citep{Feast1989, Wood1996}.

\section{Discussion and Summary}
\label{sec:disc}

We have presented spectroscopic and photometric evidence to suggest that the donor star in Sct\,X-1 is a M8-9 III type pulsating Mira, revising the classification of \citet{Kaplan2007} who favored a RSG donor. Most SyXRB donors have been previously classified with early M-type red giant donors, similar to the first classified source GX\,1+4 \citep{Davidsen1977, Chakrabarty1997} that was suggested to be at the tip of first ascent Red Giant Branch. Using an optical spectrum, \citet{Smith2012} classified the very red donor of the symbiotic fast X-ray transient XTE\,J1743-363 as a possible $>$M7 AGB star; however, the donor photometric variability remains unconstrained. CXOGBS\,J173620.2-293338 and CGCS\,5926 were also proposed as candidate SyXRBs spatially coincident with Carbon stars \citep{Masetti2011, Hynes2014}, but their low X-ray luminosity does not exclude white dwarf (WD) accretors. 

Sct\,X-1 thus represents the first confirmed Mira donor star in an X-ray binary, akin to the relatively rare class of D-type symbiotic stars hosting WDs \citep{Allen1974}. Our results are consistent with the classification scheme of \citet{Akras2019a, Akras2019b}, who classified Sct\,X-1 as a likely D-type symbiotic binary based on its IR colors. However the color-based classification is not completely reliable since they did not account for interstellar extinction, which is a substantial correction in the case of Sct\,X-1. 4U\,1954+31 remains the only confirmed Galactic RSG SyXRB \citep{Hinkle2020}, noting that the donor of CXO\,174528.79-290942.8 remains to be spectroscopically confirmed; \citet{Gottlieb2020} could not exclude the possibility of a AGB donor based on the NIR colors.

\begin{figure}
    \centering
    \includegraphics[width=\columnwidth]{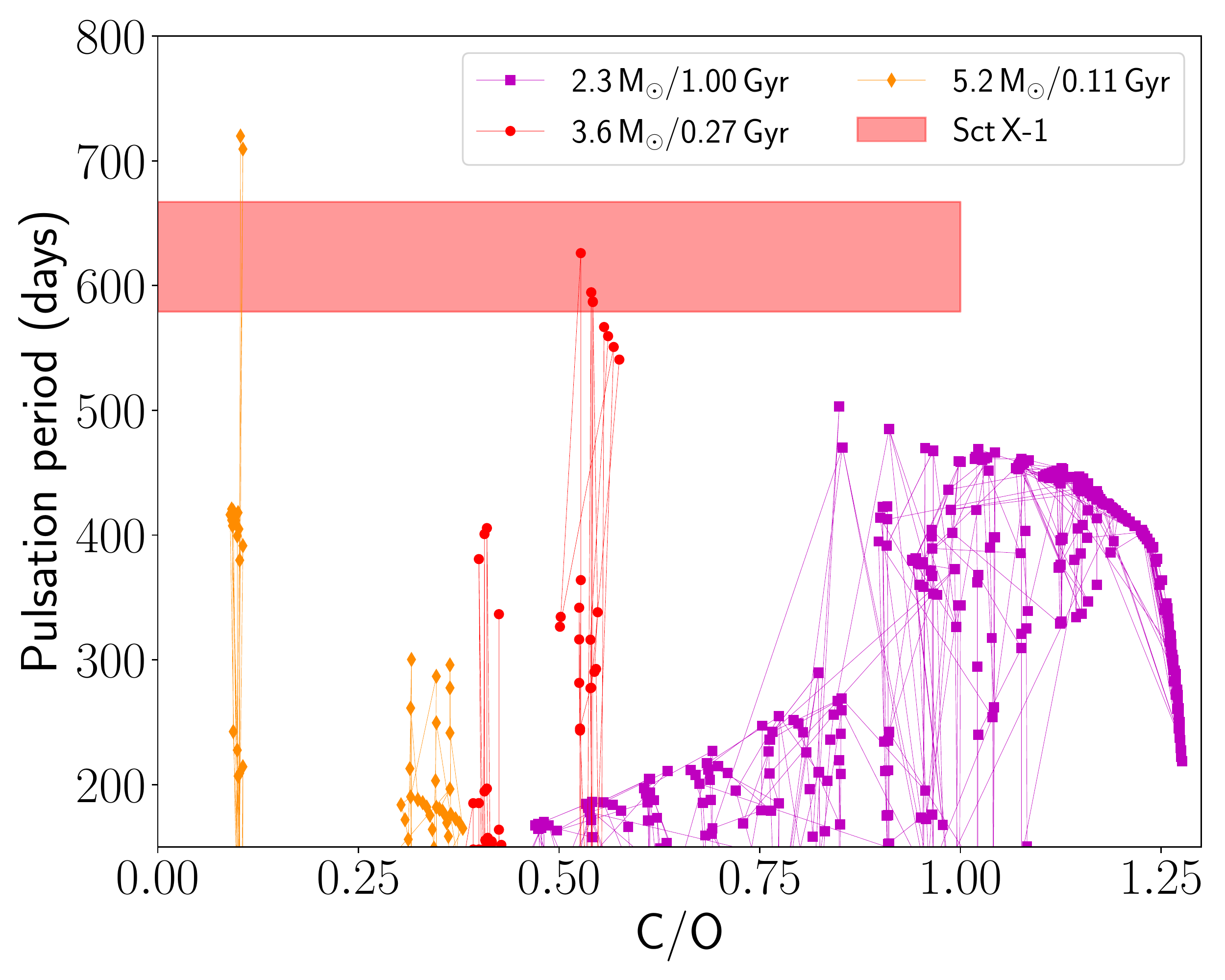}
    \caption{Comparison of the observed pulsation period and surface composition of the infrared counterpart of Sct\,X-1 (shown in the red shaded region) to evolutionary tracks of AGB stars at different initial masses for solar metallicity composition ($Z = 0.0152$). We show tracks for three masses as indicated in the legend (along with the corresponding stellar ages), with lines connecting the model epochs in increasing age sequence during the thermally pulsing phase. The period for the dominant pulsation mode at each stage is shown \citep{Trabucchi2019}.}
    \label{fig:agbmodel}
\end{figure}

Our identification of a $\approx 621$\,day pulsation period in 2MASSJ1835 makes the donor one of the longest period symbiotic Miras known in the Galaxy \citep{Gromadzki2009}. Long period Miras ($>500$\,d) are known to be relatively rare; their rarity reflects the brevity of the long period pulsation phase for lower mass AGB stars, or the rarity of higher mass AGB stars that spend most of their thermally pulsing phase at longer periods \citep{Vassiliadis1993, Marigo2017, Trabucchi2019}. We can thus constrain the evolutionary stage of the system by comparing the observed pulsation period and surface composition (O-rich; C/O$<1$) to recent AGB evolution and pulsation models \citep{Marigo2017, Trabucchi2019, Trabucchi2021}. To this end, we show in Figure \ref{fig:agbmodel}, evolutionary tracks\footnote{The tracks were obtained from the online CMD interface at \url{http://stev.oapd.inaf.it/cgi-bin/cmd_3.6}.} of surface composition and pulsation periods for different initial masses.

AGB stars with lower initial masses ($\lesssim 1.5$\,M$_\odot$) do not evolve to the observed long period \citep{Vassiliadis1993, Trabucchi2019}. While the $2.3$\,M$_\odot$ model does evolve to $P > 500$\,d, stars in the initial mass range $\approx 2.0-2.5$\,M$_\odot$ are known to evolve to Carbon Miras, inconsistent with the O-rich composition. As shown in Figure \ref{fig:agbmodel}, we find that stars in the initial mass range $\approx 3.0-5.0$\,M$_\odot$ evolve to the long observed period with an O-rich composition (due to the hot bottom burning process destroying carbon in this higher mass range; \citealt{Hofner2018}). We thus favor the scenario where the donor in Sct\,X-1 had an initial mass of $\approx 3.0-5.0$\,M$_\odot$ star, corresponding to a current binary system age of $\approx 0.1-0.3$\,Gyr. Our conclusions are consistent with the period-age relationship of Mira variables discussed in the recent work of \citet[their Figure 2 and Figure B.1]{Trabucchi2022}, and suggest that Sct\,X-1 is an intermediate mass X-ray binary.

AGB donors have been long predicted to exist in X-ray binaries based on orbits of radio pulsars (e.g. \citealt{vandenHeuvel1994}); however, known radio pulsars probe binary populations with closer orbits ($\sim$days$-$months). On the other hand, the formation of wide-orbit SyXRBs requires fine-tuned conditions. If the progenitor binary is too wide, it is likely to get disrupted during the preceding SN or never reach the high wind accretion rate required to produce symbiotic activity. If the binary is too close, the NS may merge with its companion during the late stage evolution of the initially lower mass secondary star \citep{Iben1995}. The inferred temperature ($\approx 2000$\,K) and luminosity ($\approx 10^4$\,L$_\odot$) suggest a Mira radius of $\approx 850$\,R$_\odot$. Taking the current Mira and NS mass to be $\approx 3$\,M$_\odot$ and $\approx 1.3$\,M$_\odot$ respectively, the inferred radius constrains the orbital period to be $\gtrsim 12$\,yr if the accretion is wind-fed \citep{Kaplan2007}, making Sct\,X-1 possibly the longest orbital period SyXRB known \citep{Hinkle2019}.

Population synthesis calculations \citep{Lu2012, Yungelson2019} suggest that AGB donors should be rare among wind-accreting SyXRBs. The NS spin period in Sct\,X-1 overlaps with the predicted range for AGB donors (Figure 3 in \citealt{Yungelson2019}), but our estimated orbital period is larger than predicted in their models ($\lesssim 10^3$\,days). This discrepancy is a consequence of their models predicting wind-accreting SyXRBs only in the donor's early AGB phase in closer orbit binaries. Specifically, \citet{Lu2012} suggested that AGB donors in the thermally pulsing phase should overflow their Roche lobes resulting in luminous X-ray emission ($L_X \gtrsim 10^{36}$\,erg\,s$^{-1}$) unlike known wind-fed SyXRBs. This is also inconsistent with the recent low recent X-ray luminosity of Sct\,X-1 ($L_X \lesssim 10^{35}$\,erg\,s$^{-1}$) that is suggestive of wind-fed accretion \citep{Kaplan2007} from a thermally pulsing AGB star, and hence requires a wider orbit than produced in their calculations.

Despite being one of the brightest X-ray sources known from the early days of X-ray astronomy, Sct\,X-1 demonstrates the potential of wide-field infrared time domain surveys combined with broadband infrared spectroscopy on moderate aperture telescopes to understand this population of intrinsically red and obscured sources. We note that donor classifications in SyXRBs have been commonly obtained from optical spectroscopy, while the NIR region contains the most sensitive tracers of the spectral type and luminosity class for late-type evolved stars \citep{Messineo2021}. We thus advocate for a uniform NIR spectroscopic and photometric classification of SyXRBs to constrain their true demographics. In particular, the existence of a significant population of Mira AGBs in SyXRBs would suggest the need for a larger fraction of wide-orbit SyXRB progenitors as well as small birth kicks during the NS formation to keep these systems bound after the supernova (SN) explosion. In fact, \citet{Yungelson2019} suggested that nearly 80\% of SyXRBs may arise of electron-capture SNe due to the resulting low birth kicks. Alternatively, \citet{Hinkle2006} suggested that the evolution scenario in SyXRBs creates ideal conditions for formation of NSs via Accretion Induced Collapse of WDs, making them uniquely suited to search for this predicted population.

\section*{Acknowledgements}
We thank S. Rappaport, M. Boyer and D. Kaplan for valuable discussions. Support for this work was provided by NASA through the NASA Hubble Fellowship grant \#HST-HF2-51477.001 awarded by the Space Telescope Science Institute, which is operated by the Association of Universities for Research in Astronomy, Inc., for NASA, under contract NAS5-26555. RS acknowledges grant number 12073029 from the National Science Foundation of China. We acknowledge the Visiting Astronomer Facility at the Infrared Telescope Facility, which is operated by the University of Hawaii under Cooperative Agreement no. NNX-08AE38A with the National Aeronautics and Space Administration, Science Mission Directorate, Planetary Astronomy Program. The authors wish to recognize and acknowledge the very significant cultural role and reverence that the summit of Mauna Kea has always had within the indigenous Hawaiian community. PGIR is generously funded by Caltech, Australian National University, the Mt Cuba Foundation, the Heising-Simons Foundation, the Bi-national Science Foundation. PGIR is a collaborative project among Caltech, Australian National University, University of New South Wales, Columbia University, and the Weizmann Institute of Science. The scientific results reported in this article are based in part on data obtained from the Chandra Data Archive. This research has made use of software provided by the Chandra X-ray Center (CXC) in the application packages CIAO, ChIPS, and Sherpa.

\bibliography{sample631}{}
\bibliographystyle{aasjournal}

\appendix
\section{The optical colors of Sct\,X-1}
\label{sec:app_colors}
Using follow-up observations using the MagIC instrument on the Magellan Baade telescope, \citet{Kaplan2007} reported 2MASS\,J1835 had optical fluxes of $r > 25.2$\,mag, and $i \approx 23.6$\,mag. However, the source was reported in the Gaia catalog at $G \approx 20.4$\,mag, $G_{RP} \approx 18.35$\,mag and $G_{BP} \approx 22.34$\,mag. Here, we investigate this discrepancy accounting for the different filters used in the observations. Since Sct\,X-1 is undetected at $\lambda < 1\,\mu$m and the IRTF library is limited to $\lambda > 0.7\,\mu$m, we use the optical spectrum of the M6-III Mira variable  V1334\,Sgr from the X-shooter spectral library \citep{Gonneau2020} to estimate the effects of the heavy reddening on the observed magnitudes. We simulate the extinction to Sct\,X-1 by reddening the observed spectrum of V1334\,Sgr with $E(B-V) = 6.5$ (assuming a \citealt{Cardelli1989} extinction law with $R_V = 3.1$), such that the resulting $J-K$ color of the comparison star is the same as 2MASS\,J1835. Performing photometry on the reddened spectrum using the respective filter profiles\footnote{\url{http://svo2.cab.inta-csic.es/theory/fps/}}, we find expected colors of $G_{RP} - r \approx 8.5$\,mag, $G_{RP} - i \approx 2.7$\,mag and $G_{BP} - G_{RP} \approx 10$\,mag. 

Thus, the large difference between the $G_{RP}$ and $r$ filters for such red sources explains the non-detection of 2MASS\,J1835 in the $r$-band observation of \citet{Kaplan2007}. However, the relatively blue observed color of $G_{BP} - G_{RP} \approx 4$\,mags compared to the simulated M6-III star suggests that pulsational variability is also required to explain the observations. Although the exact dates of observation are not available in Gaia EDR3, the EDR3 observations were obtained over a period of $\approx 34$\,months, spanning $\approx 1.5$ cycles of the estimated pulsation period. For the observed $J$-band amplitude of $\approx 3.5$\,mags, the estimated amplitudes in optical filters would be $\gtrsim 6$\,mags given the typical wavelength dependence of amplitude in O-rich Miras \citep{Iwanek2021}. Gaia\,EDR3 reports \texttt{phot\_bp\_n\_obs} = 2 and \texttt{phot\_rp\_n\_obs} = 9 for 2MASS\,J1835, suggesting that the source was indeed detected in more epochs in the redder filter where the source is brighter and has smaller flux amplitudes. On the other hand, the $G_{BP}$ flux is likely dominated by observation epochs near the pulsation maximum where Miras are bluest in color \citep{Reid2002}. Thus, the reported Gaia EDR3 colors would be closer to the bluest color of the source near maximum light.

\end{document}